\pdfoutput=1
\documentclass[a4paper]{article}
\usepackage{authblk}
\usepackage[version=3]{mhchem}
\usepackage{graphicx}
\usepackage{subfig}
\usepackage{multirow}
\usepackage{longtable}
\usepackage{achemso}
\usepackage{array,booktabs}
\usepackage{float}
\usepackage{caption}
\usepackage{color}
\usepackage{enumerate}
\usepackage{amsmath}
\usepackage[none]{hyphenat}

\begin{document}
\title{Linear and Nonlinear Optical Properties of Graphene Quantum Dots: A Computational Study.}

\author[1]{Sharma SRKC Yamijala}
\author[2]{Madhuri Mukhopadhyay}
\author[2,3,*]{Swapan K Pati \thanks{pati@jncasr.ac.in}}
\affil[1]{Chemistry and Physics of Materials Unit, Jawaharlal Nehru Centre for Advanced Scientific Research, Bangalore 560064, India.}
\affil[2]{Theoretical Sciences Unit, Jawaharlal Nehru Centre for Advanced Scientific Research, Bangalore 560064, India.}
\affil[3]{New Chemistry Unit, Jawaharlal Nehru Centre for Advanced Scientific Research, Bangalore 560064, India.}
\affil[*]{Corresponding author}
\date{}

\maketitle


\begin{abstract}

Due to the advantage of tunability via size, shape, doping and relatively low
level of loss and high extent of spatial confinement, graphene quantum dots
(GQDs) are emerging as an effective way to control light by molecular
engineering. The collective excitation in GQDs shows both high energy plasmon
frequency along with frequencies in the terahertz (THz) region making these
systems powerful materials for photonic technologies. Here, we report a
systematic study of the linear and nonlinear optical properties of large
varieties of GQDs ( $\sim$ 400 systems) in size and topology utilizing the
strengths of both semiempirical and first-principles methods. Our detailed
study shows how the spectral shift and trends in the optical nonlinearity of
GQDs depends on their structure, size and shape.  Among the circular,
triangular, stripe, and random shaped GQDs, we find that GQDs with inequivalent
sublattice atoms always possess lower HOMO-LUMO gap, broadband absorption and
high nonlinear optical coefficients.  Also, we find that for majority of the
GQDs with interesting linear and nonlinear optical properties have zigzag
edges, although reverse is not always true.  We strongly believe that our
findings can act as guidelines to design GQDs in optical parametric
oscillators, higher harmonic generators and optical modulators.

%
%

\end{abstract}



\section{Introduction}

	Materials with broadband absorption (BBA) and emission, that is,
covering ultraviolet, visible, and near-infrared regions of the solar spectrum,
have important applications in photodetectors, broadband modulators,
bioimaging, solar cells and so forth.  \cite{Zheng2014,Sun2014, Kim2014,
Sobon2013, Urich2011, Sun2010} Moreover, if the materials with the broadband
absorption also shows optical nonlinearity they can be very useful in
applications involving optical parametric oscillation, high harmonic
generation, \cite{ Sørngard2013, Hong2013} Kerr effect \cite{
Shen2014,Shimano2013} and multiphoton imaging. \cite{ Yang2011} Thus, finding
novel materials with both broadband absorption and optical nonlinear activity
is of great interest.

	Group IV-VI quantum dots like CdSe, PbSe, CdS, HgTe, ZnSe, etc. have
already been there in variety of applications involving light emitting diodes,
bio-imaging, solar cells, and so forth, because of their tunable absorption and
specific optical nonlinear activity.  \cite{ Mashford2013, Klimov2000,
Choi2014, Tian2010, Somers2007, Lad2007} Materials prepared from high band gap
semiconductors like ZnS, ZnSe, GaN, and AlN possess ultraviolet optical
activity whereas CdS, rare earth doped GaN materials exhibit near IR
activities. \cite{Shavel2004, Dennis2012} Although, tuning the size of a
quantum dot can vary its active optical range, it cannot give the whole range
altogether (i.e.  simultaneously UV-VIS and IR range activity). To this end,
GQDs and modified GQDs seems to be promising materials for such optical
activities \cite{Yamijala2013,Rieger2010}. Together with their higher
photostability, bio-compatibility and low cost preparation, GQDs may act as a
substitute for the toxic IV-VI  group quantum dots.
  
	GQDs are the confined graphene materials available in various
topologies \cite{Silva2010, Kuc2010, Rieger2010} and  graphene is a layered
sp$^{2}$-bonded carbon material in honeycomb lattice.  Graphene with its zero
band gap has a limitation to its applications in optoelectronics due to its
zero optical emission. On the other hand, GQDs exhibit a broadband absorption
and they have emerged as attractive fluorescence materials in the ultraviolet,
visible and even in infrared regions.  \cite{Rieger2010,Yamijala2013,Sk2014}
During recent years, there has been a lot of research on the broadband activity
of GQDs of different sizes, shapes and functionalities through both experiment
and theory.  \cite{Fang2013, Yan2013, Tang2014, Riesen2014} Also, there is a
progress in identifying the shape and size dependent nonlinear activity of
GQDs.  \cite{Zhou2011,Yoneda2009, Yoneda2012, Yoneda2012a} 

	Considering these studies into account, here, we have performed a
systematic computational study on the linear and nonlinear optical (NLO)
properties of hydrogen passivated GQDs (hence, may also be termed as
polyaromatic hydrocarbons (PAHs)) of various sizes, shapes, edge structures and
so forth. 
After careful analysis on these GQDs ($\sim$ 20) with simultaneous BBA and high
NLO coefficients, we find that the necessary and sufficient condition for
possessing such multifunctionality is the presence of inequivalent sublattice
atoms.  Also, we find that majority of the GQDs with only zigzag edges possess
this multifunctionality. Additionally, we find that some of these GQDs show
fascinating 1st hyperpolarizabilities ($\sim$ 10$^{3}$-10$^{5}$ times larger
than the traditional NLO compounds [like p-nitroaniline etc]). In the
following, first we have described how we have modeled our systems and then we
have given the details of our computations. Next, we have compared the results
from our semiempirical calculations on structural stability and electronic
properties with the earlier studies and then we have presented our results on
linear and nonlinear optical properties.  Finally, we have presented the
results from first principles calculations on the systems, followed by the
conclusions and possible extensions to the present work.  

\begin{figure}
\center\includegraphics[scale=0.25]{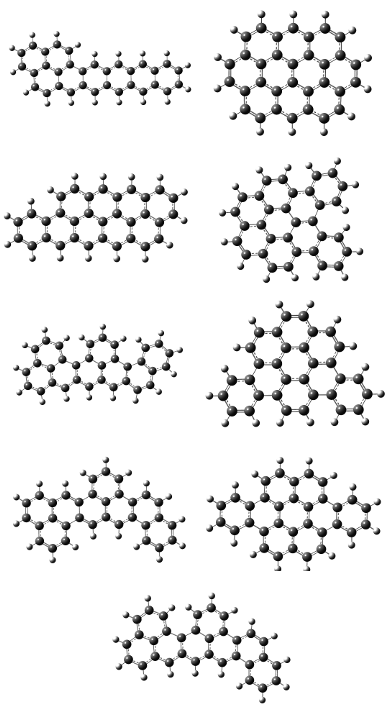}
\caption{Some random shaped GQDs with 32 carbon atoms (C32)}
\end{figure}


\section{Modeling and Computational Details}

	As the number of varieties of GQDs which can be generated from graphene
are huge, following Kuc et. al.\cite{Kuc2010} we have considered $\sim$ 400
structures, based on their size, shape, edge etc. As the hydrogen passivated
GQDs have been shown to be more stable than the GQDs with bare edges, we have
only considered the former ones throughout our study. As in ref \cite{Kuc2010},
we have categorized our GQDs as circular (F) or triangular (T) or stripes (i.e.
nanoribbons) (S) depending on their shape and zigzag (z) or armchair (a)
depending on their edges. Thus, F$_{a}$ (T$_{z}$) represents circular
(triangular) GQDs with armchair (zigzag) edges. All other GQDs which don't fit
in these categories mainly represent the different possible conformers of a GQD
with particular number of carbon atoms and we refer them as random shaped GQDs.
We identify these random shaped GQDs with their carbon atom numbers such as
C22, C28, C74 etc. In Figure 1 we have shown typical examples of random shaped
GQDs. 


	All the structural optimizations have been performed using
self-consistent charge (SCC) density functional tight-binding (DFTB) theory
\cite{Elstner1998} within third order expansion of the energy (DFTB3)
\cite{Yang2007} and with 3ob parameter set, \cite{Gaus2013} as implemented in
DFTB+ package.\cite {Aradi2007} DFTB level of theory is used mainly due to the
large number of systems ($\sim$ 400) considered in this study as well as its
ability to give trends in band-gaps, energies etc. which are comparable to the
ones given by DFT, especially for carbon related materials, even with different
edges, defects and so forth.\cite{Zobelli2012, Enyashin2011} Geometry
optimizations have been performed using conjugate gradient method and systems
are considered to be optimized only when forces on all the atoms are less than
0.0001 Ha/Bohr. As the systems are zero-dimensional, we have performed the
$\Gamma$-point calculations. For those systems whose energy levels near the
Fermi-level are almost degenerate, we have increased the electronic temperature
to 100 K to avoid any convergence issues.
  
	Linear optical properties of all the compounds have been computed at
the semi-empirical ZINDO/S level of theory as implemented in g09 software
package. \cite{g09} ZINDO/S has been proved to be very successful especially in
predicting the optical properties of systems containing C, N, O, H atoms like
polyaromatic hydrocarbon compounds, \cite{Ona-Ruales2014, Cocchi2014}
chlorophylls \cite{Linnanto2000} etc. \cite{Voityuk2013} At semi-empirical
level, nonlinear optical (NLO) properties of all compounds have been calculated
using MOPAC 2012 program package. \cite{MOPAC2012, Maia2012} All the first
principles calculations for the linear (at time dependent density functional
theory (TDDFT) level) and nonlinear optical properties have been performed
using g09. Long range corrected (CAM-B3LYP) exchange correlation functional has
been used in conjunction with 6-31+g(d) basis set for all the calculations. A
minimum of first 12 lowest excited states have been considered in all the
studies. 
GaussSum-2.2.6.1 \cite{OBoyle2008} is used to plot the absorption spectra and a
broadening of 0.333 eV has been used. To ensure the reliability of our
calculations, we have compared our semi-empirical and first principles results
on the linear and nonlinear optical properties of p-nitroaniline with its
reported experimental values. We find that both the results are in close
agreement (see Table S1 of supporting information (SI)).  


\section{Results and Discussion}

\subsection{Energetic Stability and Electronic Properties:}

	All the GQDs considered in this study are found to be thermodynamically
stable, that is, they have negative formation energy, E$_{Form}$ = E$_{tot}$ -
N$_H$*E$_H$ - N$_C$*E$_C$, where E$_{tot}$, E$_H$ and E$_C$ are the total
energy of the system, energy of the hydrogen atom in a H$_{2}$ molecule (i.e.
E$_{H2}$/2) and energy of the carbon atom in a graphene lattice (i.e.
E$_{Graph}$/N$_{C}$), respectively.  Here, N$_{C}$ and N$_{H}$ are the number
of carbon and hydrogen atoms in the system. At DFTB3 level of theory, we find
E$_{H}$ and E$_{C}$ to be -9.123 and -44.291 eV, respectively.  A plot of
formation energy per atom vs  N$_{H}$/(N$_{H}$+N$_{C}$) of all the systems is
given in Fig. 2a. Clearly, there is a near linear relationship between the
formation energy per atom and the number of edge atoms in all the systems
(notice the linear fit in Fig. 2a), that is, system with lesser number of edge
atoms is easier to form and vice-versa, as expected.
\cite{Yamijala2013,Yamijala2014,Bandyopadhyay2013}  Similar results have been
observed in some of the earlier studies on GQDs and PAHs 
\cite{Fthenakis2013, Kuc2010}  In agreement with these previous studies, we
also find that among the different GQD shapes studied here, circular GQDs are
the most stable ones and ribbon like GQDs are the least stable. All other GQDs'
(triangular, random etc.) stability fall in between these two types of GQDs
[see Fig. 2a]. The reason for such a trend is again due to the less number of
edge atoms in circular GQDs than in other GQDs considered in this study, as
evident from the x-axis of Fig. 2a.  Recent molecular dynamics simulations have
also shown that among the different GQDs, circular and triangular GQDs with
zigzag edges as the most stable ones till $\sim$ 4000 K.  \cite{Silva2010}

	Next, the energies of HOMO, LUMO and their difference (i.e. HOMO-LUMO
gap (HLG)) of all the GQDs are plotted in Fig. 2b as a function of number of
carbon atoms. The calculated HLG values are mainly in the range of $\sim$ 0-3
eV. Also, from Fig. 2b and 2c, it can be observed that for a particular
N$_{C}$, one can tune the HLG from $\sim$ 0-3 eV depending on the shape and
edges of the GQD. Also it should be noted that such tuning is possible even for
the systems with N$_{C}$ between 20 to 50. In fact, synthesis of GQDs
(actually, PAHs) of different sizes have already been carried out.
\cite{Rieger2010} From Fig. 2c, it can be noticed that HLG of the systems with
zigzag edges converge rapidly to zero (reaching the semi-metallicity of
graphene) than the armchair ones, irrespective of the shapes and the calculated
trend of convergence is T$_{z}$-GQDs $>$ S$_{z}$-GQDs $>$ F$_{z}$-GQDs $>$
T$_{a}$-GQDs $\sim$ F$_{a}$-GQDs $>$ S$_{a}$-GQDs. As HLG reflects the kinetic
stability of a system, the above trends suggest that kinetic stability will be
highest for S$_{a}$-GQDs and least for T$_{z}$-GQDs and S$_{z}$-GQDs. As
suggested by the Clar’s rule, \cite{Rieger2010} higher kinetic stability of
S$_{a}$-GQDs, compared to the other structures is due to the presence of larger
number resonant sextets in these structures.  Similar reasons are also known
for the lesser stability of zigzag edged structures compared to the armchair
ones. \cite{Kuc2010} One may also notice that the HLG of ``S$_{z}$ and
T$_{z}$", ``T$_{a}$ and F$_{a}$"-GQDs follows similar trend as N$_{C}$
increases (for N$_{C}$ $>$ 60) as has also been observed in some of the recent
studies.  \cite{Silva2010} Finally, as the HLG of these GQDs are tunable over a
wide range and as HLG can be used as a rough estimate for the optical
gap,\cite{Ona-Ruales2014} one may immediately expect that the optical
properties of these GQDs can also be tuned over a wide range and the results of
the respective calculations are given below.

\begin{figure}
\center\includegraphics[scale=0.2]{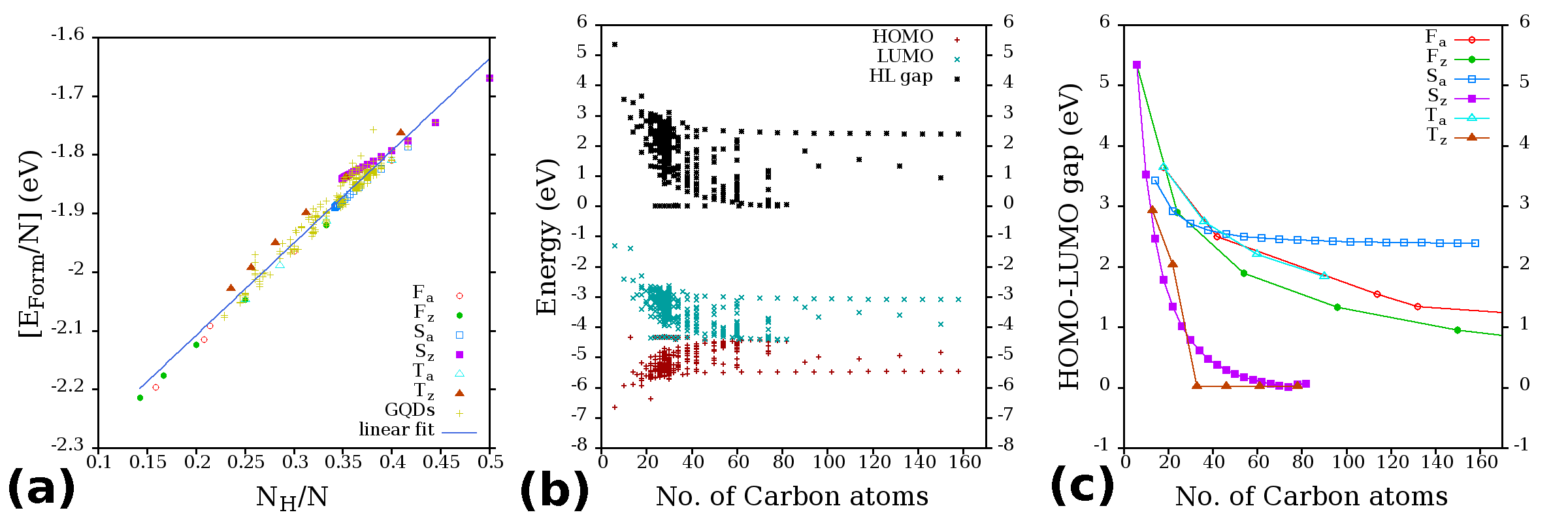}
\caption{(a) A plot of E$_{Form}$ per atom versus number of edge atoms to the
total number of atoms (N) of all the GQDs. Straight line shows the linear fit.
(b) Energies of HOMO, LUMO and the HOMO-LUMO gap (HLG) of all the GQDs and (c)
Changes in the HLG with size for different shaped GQDs.  Symbols T, F, S
represents triangular, circular and striped GQDs. Subscripts a and z represents
armchair and zigzag edges. See the ``Modeling" for further details.}
\end{figure}


\subsection{Optical Properties:} 


	First, we present the optical absorption of all the systems calculated
at the ZINDO/S level of theory. Here, we have analyzed only the 20 low energy
singlet excitations from the ZINDO/S results. Absorption spectra of PAHs mainly
consists of 3 bands, namely, alpha ($\alpha$), beta ($\beta$) and para
($\textit{p}$), out of which the most intense ones being $\beta$ and
$\textit{p}$-bands (notations are according to Clar's rule \cite {Rieger2010},
where \textit{p}($\beta$)-bands corresponds to the bands at higher (lower)
wavelengths). Interestingly, in a very recent study \cite{Ona-Ruales2014} it
has been concluded that ZINDO/S is good at predicting the most intense
$\textit{p}$ and $\beta$ bands of all C$_{32}$H$_{16}$ benzenoid PAHs.
Considering these facts, first we have plotted the histograms of ``wavelengths
corresponding to the most intense $\textit{p}$-bands ($\textit{p}_{max}$) and
$\beta$-bands ($\beta_{max}$)", respectively, in Figs.  3a and 3b and the
corresponding oscillator strengths (OS) histograms in Figs. S1a and S1b. From
these figures it can be noticed that, majority of the systems have their
$\beta_{max}$ and $\textit{p}_{max}$ in the UV-VIS region (200-760 nm) and the
oscillator strength of $\beta_{max}$ ($\textit{p}_{max}$) is almost always (for
majority of structures) $>$ 0.5 (0.1).  Thus, majority of GQDs considered in
this study absorb strongly in the UV-VIS region (in particular, their
$\beta_{max}$ ($\textit{p}_{max}$) is located in the region between 250-450
(300-700) nm).  

\begin{figure}
\center\includegraphics[scale=0.2]{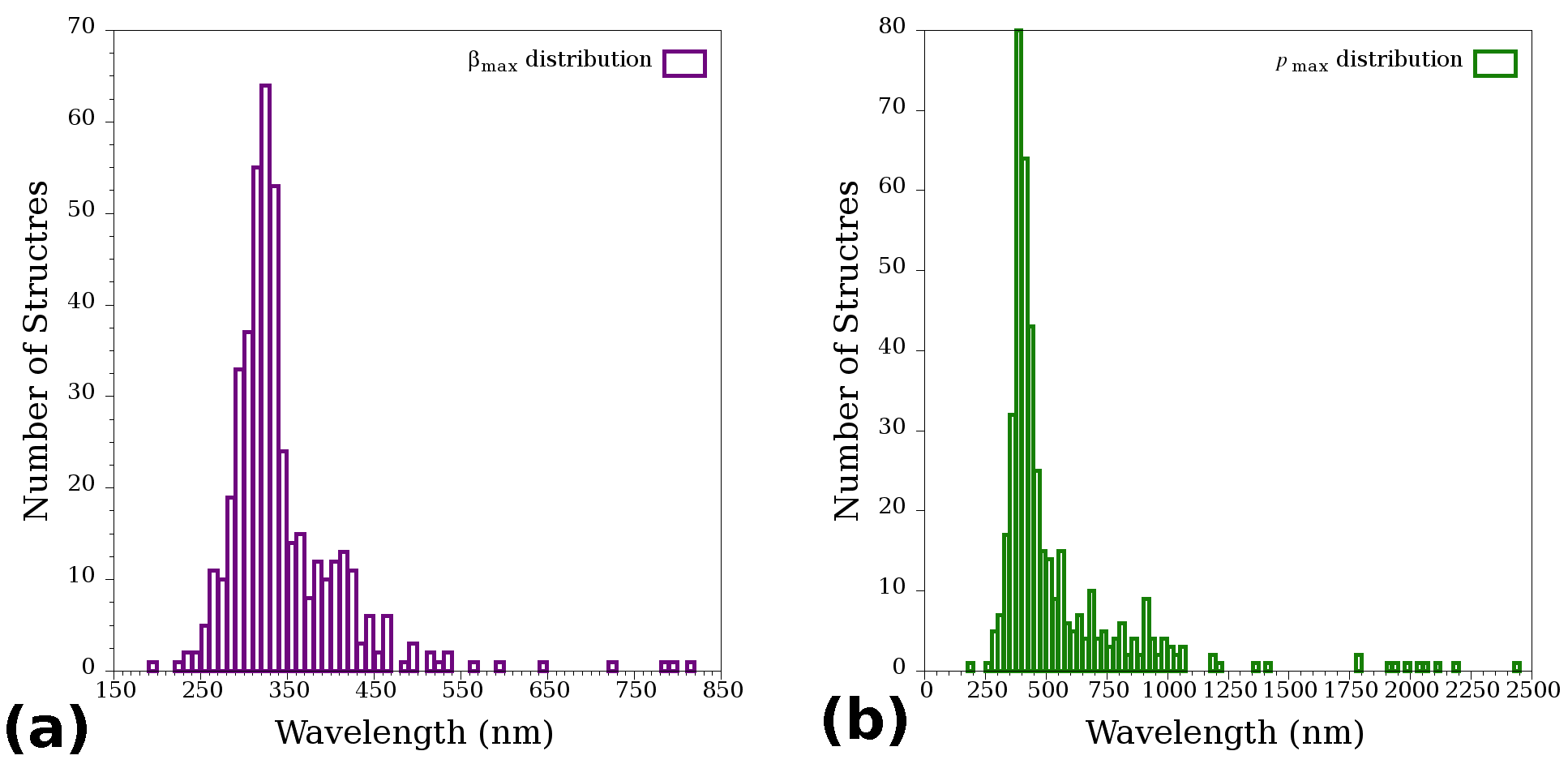}
\caption{Histograms of wavelengths corresponding to (a) $\beta_{max}$, (b)
$\textit{p}_{max}$ excitation in all GQDs. $\textit{p}_{max}$ excitations above
2500 nm have been omitted for clarity.}
\end{figure}

	However, interestingly, we find $\sim$ 70 GQDs whose $\textit{p}_{max}$
is in IR-region ($>$ 760 nm). Materials absorbing in IR region are of great
interest in the preparation of solar cells because half of the solar energy
received by earth is in IR radiation range and most of the  present day solar
cells do not utilize this energy region. \cite{Yamijala2013} Thus, knowing the
reason for the IR-activity of these GQDs will be of great use and for this we
have analyzed their $\textit{p}_{max}$ transition.  In Table S1, we have given
the calculated wavelength, OS and the major contributions of the molecular
orbitals corresponding to the $\textit{p}_{max}$ transition for all the
IR-active GQDs.  Clearly, $\textit{p}_{max}$ transition always has the major
contributions from excitations involving the frontier orbitals (that is,
HOMO-1, HOMO, LUMO and LUMO+1), especially from HOMO and LUMO. Thus, the
changes in these frontier MOs lead to changes in the $\textit{p}_{max}$
transition.  Also, some of the earlier studies on PAHs have found that HLG of
these systems is almost equal to the energy corresponding to the
$\textit{p}_{max}$ transition (see Ref \cite{Ona-Ruales2014} and references
there in). 

	For a few of these GQDs, we find HLG to be very small. In general,
small HLGs occur either due to extended delocalization (as in conjugated carbon
chains) or if there exists lesser number of resonant sextets (according to
Clar's rule \cite{Ona-Ruales2014,Rieger2010,Kuc2010}). In our case, however,
the very small HLGs are seen due to completely different reasons. If we look at
the structures of these GQDs closely, we find that, they don't have same number
of sublattice atoms (i.e. N$_{A}$ - N$_{B}$ $\neq$ 0). In fact, in all the
random shaped GQDs, we find there exists two additional sublattice atoms of one
type (i.e. $|$ N$_{A}$ - N$_{B}$ $|$ = 2).  $\textit{p}_z$ orbitals of these
additional atoms remains as non-bonding orbitals and appear at the zero of
energy (i.e. at the Fermi-level) in the energy level diagram. If there were no
interactions (as in tight-binding calculations), both of these levels would be
degenerate and would appear exactly at the zero of energy.  (similar to what
has been observed in triangular GQDs \cite{Fernandez-Rossier2007}). However,
because of interaction terms in ZINDO/S Hamiltonian, we find the two levels to
appear above and below the zero of energy with a very low energy gap (few meV).
Interestingly, these two levels have opposite parity, due to which the
transition dipole moment between the two become non-zero. Thus, these two
levels give rise to optical transition with a finite oscillator strength (OS).
Since, the energy gap between these two levels is too small, the optical
absorption appears in IR-region. 

	As an example, in Figs. 4a-4d we have given four structural isomers
(here after, addressed as 4a, 4b, 4c and 4d, respectively) of C$_{32}$H$_{18}$,
where only 4b and 4d have the sublattice imbalance. As explained, only for 4b
and 4d, we find $\textit{p}_{max}$ in IR-region ($>$ 2000 nm) but not for 4a
and 4c. In Fig. 4, we have also given the conjugation and isosurface plots of
HOMO for these GQDs. As can be seen, because of sublattice imbalance, the
conjugation in 4b and 4d GQDs is not continuous and there are ``conjugation
breaks", which are clear demonstration of solitonic structure (conjugated
system), domain walls (seen in ferromagnetic metal blocks). The main point is
that, these defect states are intrinsic in these GQDs and these have not been
externally induced.

\begin{figure}
\center\includegraphics[scale=0.2]{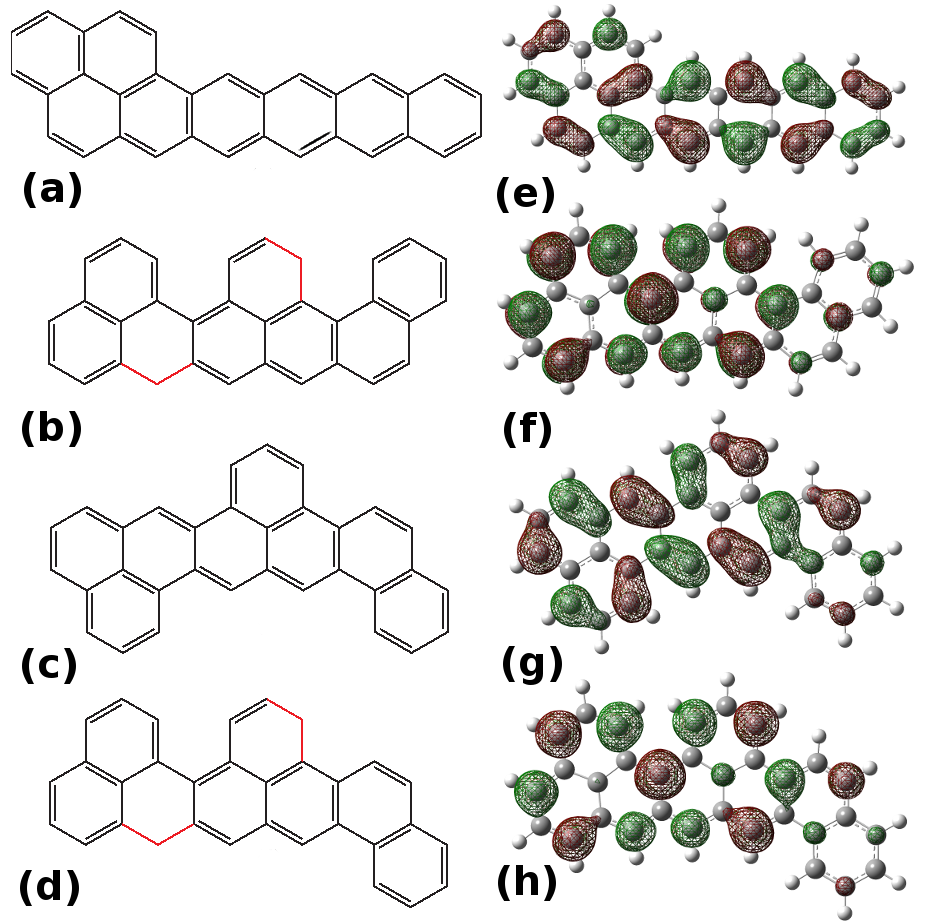}
\caption{Schematic diagrams of four structural isomers of C$_{32}$H$_{18}$ GQD
are given in (a)-(d) and their HOMO isosurfaces are given in (e)-(h),
respectively. Iso-value of 0.02 e/\AA$^{3}$ is used for all the plots.}
\end{figure}

	IR-activity of GQDs which absorb below 2500 nm is mainly due to the
zigzag edge nature of these GQDs, which lowers their HLG. For example, it is
well known that the polyacenes have the lowest HLG among the various PAHs
\cite{Rieger2010} and $\textit{p}_{max}$ of hexacene (6 fused benzne rings)
itself is 750 nm. Also, through TDDFT calculations, previously our group has
shown that $\textit{p}_{max}$ of rectangular GQDs is $\sim$ 1900 nm.
\cite{Yamijala2013} Inspecting the structures of GQDs which absorb in the
region of 760-2500 nm, we find that all these GQDs have either polyacene type
structure or rectangular type structure, with some of their edges being
substituted with ethene, propene, cis-1,3-dibutene etc. (see Fig. S3) Also, it
is important to mention that HOMO of all the IR-active structures is different
from that of the non-IR-active structures. 

	Frontier MOs of IR-active GQDs have larger number of nodes, and hence,
look like the collection of p$_{z}$ orbitals on individual carbon atoms without
overlap (for example, see Figs. 4f and 4h). The reverse is true for the
non-IR-active GQDs (see Figs. (4e, 4g)). Presence of large number of nodes
destabilizes HOMO compared to its structural isomers with less number of nodes,
and hence, lesser HLG and IR-activity. As an example of the above mentioned
observations, we have given absorption spectra and isosurfaces of HOMO of C32
and C74 GQDs in Figs. S2-S4.  Finally, as the OS of $\textit{p}_{max}$ peak for
majority of these GQDs is $>$ 0.5 and as OS of $\beta_{max}$ peak is almost
always found to $>$ 0.5, we find that these GQDs can have broad band absorption
(BBA), as predicted earlier for the rectangular GQDs. \cite{Yamijala2013} BBA
of these GQDs can also be seen in Figs. S2 and S4. To conclude all the above
results, we find that GQDs with inequivalent sublattice atoms or GQDs with
rectangular or stripe shapes can absorb in IR-region and they may be suitable
candidates for BBA.

\begin{figure}
\center\includegraphics[scale=0.2]{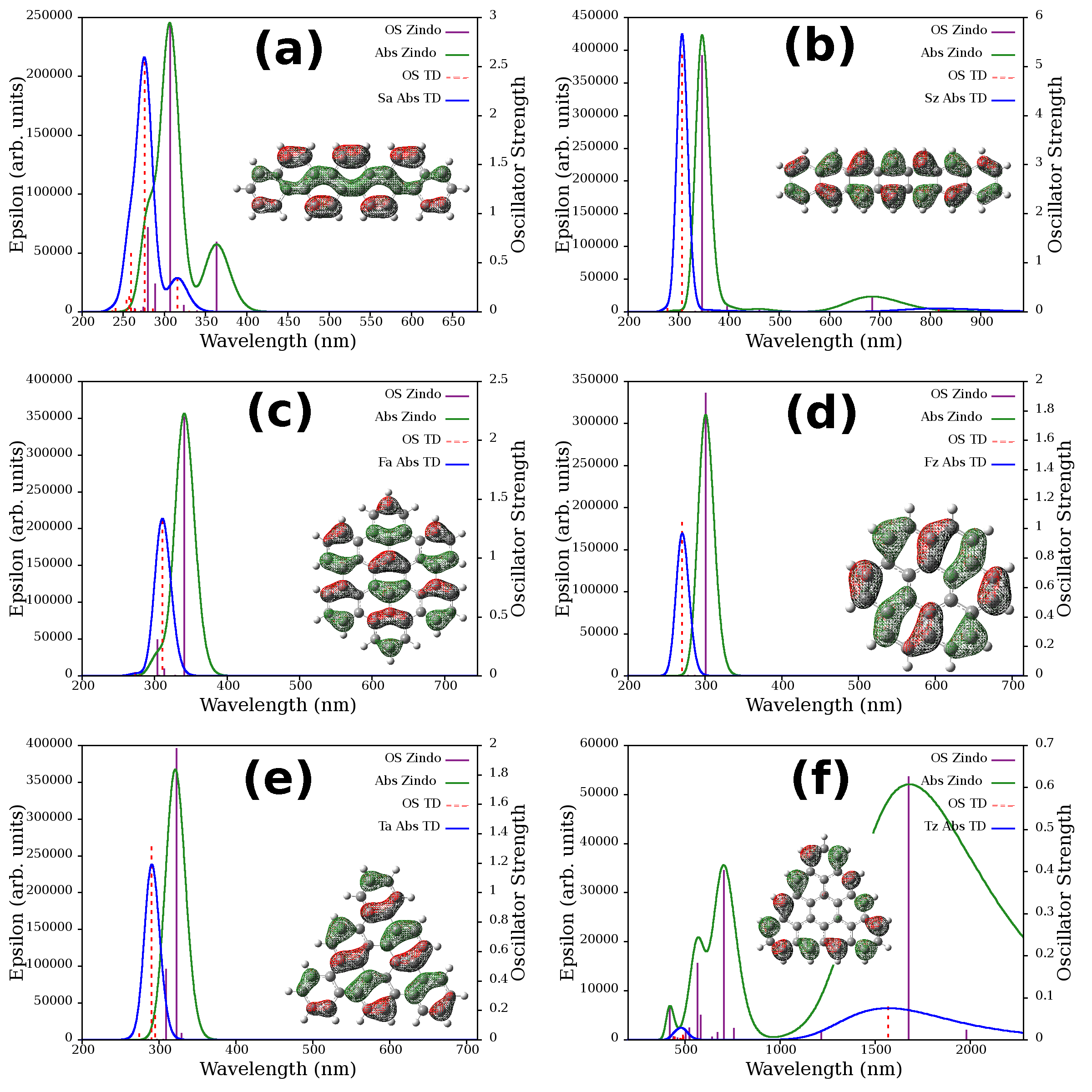}
\caption{Absorption profiles of GQDs of various shapes calculated at both
ZINDO/S level of theory and using TDDFT at CAMB3LYP/6-31+g(d) level of theory.
Insets in each figure show the isosurface of the HOMO of that GQD calculated
using TDDFT. (a)-(f) represents S$_{a}$, S$_{z}$, F$_{a}$, F$_{z}$, T$_{a}$ and
T$_{z}$ GQDs, respectively. Iso-value of 0.02 e/\AA$^{3}$ is used for all the
plots.}
\end{figure}

	To put the results obtained from the ZINDO/S method in a solid footing,
we have performed TDDFT calculations at CAMB3LYP/6-31+g(d) level of theory on a
few GQDs. First, we will present our results on GQDs of various shapes. In
Figs. 5a-5f, we have given the absorption spectra of S$_{a}$, S$_{z}$, F$_{a}$,
F$_{z}$, T$_{a}$ and T$_{z}$, respectively, calculated at both ZINDO/S and
TDDFT levels of theory along with the iso-surfaces of their HOMO (only from
TDDFT). Clearly, absorption profiles of both the methods compares farily well,
although OS values predicted by ZINDO/S are higher than that of TDDFT. Also,
$\lambda_{max}$ predicted by ZINDO/S is consistently red-shifted compared to
the TDDFT predicted values. Consistent with the previous arguments on the
isosurface of HOMO (calculated using ZINDO/S), even with TDDFT  we find larger
number of nodes (see Fig. 5f) in the HOMO if the GQD has IR-activity and it has
more overlapping character if the GQD is not IR-active (see Fig. 5a-5e). Also,
we find that the character of the $\textit{p}_{max}$ excitation (i.e. MOs
involved in the excitation) are similar in both the methods. Importantly, we
find that GQDs whose HOMO is mainly localized on the edge atoms (as in S$_{z}$
and T$_{z}$) and whose $\textit{p}_{max}$ excitation has major contribution
from HOMO to LUMO, are IR-active. From Fig. 5, one may also infer that the
presence of zigzag edges is only a necessary, but not a sufficient condition
(example being the F$_{z}$ GQDs) for the IR-absorption. Finally, to see the
effect of inequivalent sublattice atoms on the IR-activity, we have considered
five C28-GQDs with N$_{A}$ - N$_{B}$ = 2, and we find all of them to be IR
active, again consistent with the ZINDO/S results (see Table S2). Thus, we find
that, results of ZINDO/S and TDDFT are consistent and compare well for the GQDs
considered in this study.


\subsection{Nonlinear Optical Properties:}

\begin{figure}
\center\includegraphics[scale=0.2]{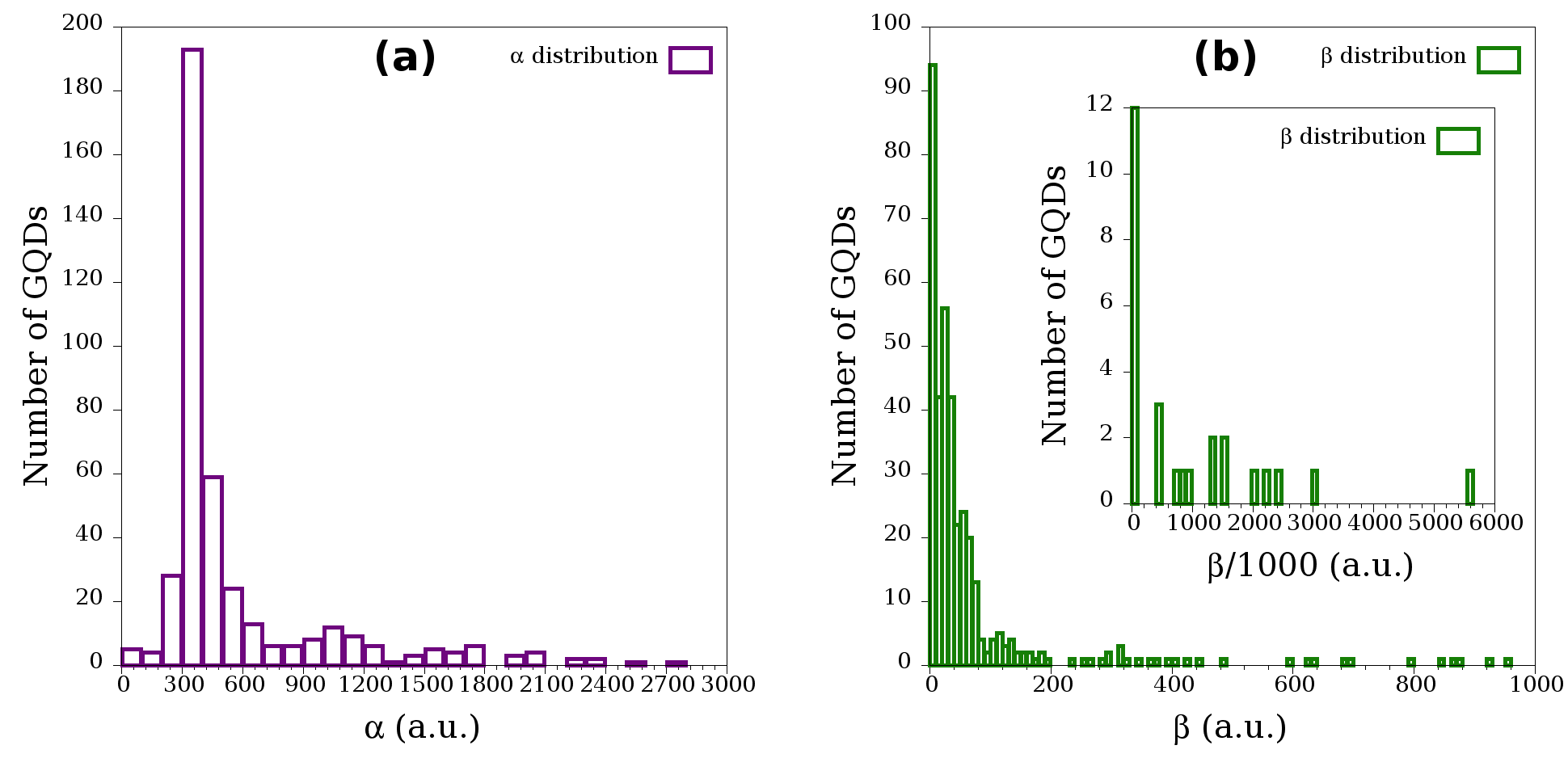}
\caption{Histograms of average values of (a) polarizability, $\alpha$ and (b) first
hyperpolarizability, $\beta$ of all GQDs.}
\end{figure}

	In this subsection, we present the linear polarizability ($\alpha$) and
first hyperpolarizability ($\beta$) of all the GQDs calculated using the finite
field approach as implemented in the MOPAC and g09 packages. Expressions for
the dipole moment and the energy of a molecule interacting with an external
electric field are given by Eqns. 1 and 2 \cite{Kurtz1990}.
\begin{eqnarray}
\mu_{i} = \mu_{0} + \alpha_{ij}F_{j} + \frac{1}{2}\beta_{ijk}F_{j}F_{k} +
\frac{1}{6}\gamma_{ijkl}F_{j}F_{k}F_{l} + \ldots \\ 
E(F) = E(0) - \mu_{i}F_{i} - \frac{1}{2!}\alpha_{ij}F_{i}F_{j} -
\frac{1}{3!}\beta_{ijk}F_{j}F_{k} - \frac{1}{4!}\gamma_{ijkl}F_{j}F_{k}F_{l} -
\ldots 
\end{eqnarray}
where, $\mu_{0}$ is the permanent dipole moment, $\alpha_{ij}$, $\beta_{ijk}$
and $\gamma_{ijkl}$ are the linear polarizability, 1$^{st}$ and 2$^{nd}$
hyperpolarizability tensor elements, respectively. Also, for a molecule, the
avergae values of above quantities ($\mu_{av}$ etc.) are defined as
\begin{eqnarray}
\mu_{av} = (\mu_{x}^{2} + \mu_{y}^{2} + \mu_{z}^{2})^{1/2} \\ 
\alpha_{av} = \frac{1}{3}(\alpha_{xx} + \alpha_{yy} + \alpha_{zz}) \\ 
\beta_{av} = (\beta_{x}^{2} + \beta_{y}^{2} + \beta_{z}^{2})^{1/2} 
\end{eqnarray}
where,
\begin{eqnarray}
\beta_{i} = \frac{3}{5}(\beta_{iii} + \beta_{ijj} + \beta_{ikk}),  i, j, k = x,
y, z
\end{eqnarray}

	In Figs. 6a and 6b, we have plotted the distribution of average
$\alpha$ and $\beta$ values for all the GQDs at static field. Similar to
absorption profiles, majority of the GQDs' $\alpha$ and $\beta$ values are
confined to a small region. For these majority GQDs, we find that the $\alpha$
and $\beta$ values are in the range of 250$-$700 a.u. ($\sim$ 40$-$100
${\AA^{3}}$) and 1$-$200 a.u.  (10$^{-32}$$-$10$^{-30}$ esu), respectively.
Compared to the $\alpha$ and $\beta$ values of para-nitroaniline (16.346 a.u.
and 978.21 a.u., respectively), it is nice to notice that majority of the GQDs
already have high polarizability and reasonable hyperpolarizabilities.
Importantly, we find that several GQDs possess $\alpha$ and $\beta$ values
which are orders of magnitude greater than that of para-nitroaniline (see Fig
6a and inset of Fig 6b).


	In general, both linear polarizability and first order
hyperpolarizabilities have an inverse relationship with the energy gap between
the states involved in the polarization, and are directly proportional to the
transition moment.  Thus, we can expect an increase in $\alpha$ and $\beta$ if
the ground and excited states are closely spaced or the transition moment
between the states is high or both. From the above reasoning, one can also
infer that GQDs with low HLG and whose 1$^{st}$ excited state has major
contribution from HOMO to LUMO transition should give higher $\alpha$ and
$\beta$ values.  Indeed, we find that all the GQDs which are IR-active also
have high $\alpha$ and $\beta$ (except the GQDs with inversion symmetry)
values, that is, above the range of 250$-$700 a.u.  and 1$-$200 a.u.,
respectively.  Also, we find that some of the GQDs with zigzag edges, like
F$_{z}$, which are not IR-active but have very high oscillator strength for the
$\textit{p}_{max}$ (see Fig. 5d) excitation also show higher $\alpha$ values.
However, to the presence of inversion symmetry, such GQDs do not have higher
$\beta$ values. Finally, we again find that trends in our results from
semiempirical calculations compares farily well with that of first-principle
calculations (see Table S4). Based on all the above results, we conjecture that
GQDs with very low HLGs can have both broad band absorption and nonlinear
optical activity, and hence, are potential candidates for optoelectronic
devices.

\section{Conclusions}

We have performed a systematic study on the GQDs of various sizes, shapes and
edges to explore their linear and nonlinear optical properties. First, we find
the formation energies of GQDs have a near linear dependence on their number of
edge atoms and HOMO-LUMO gaps of a GQD with a particular number of carbon atoms
can be tuned from $\sim$ 0-3 eV depending on its shape and edge nature. Trends
in the HLG can be understood based on the Clar's rule of aromatic sextets for
majority of the sytems.  Extremely low HLGs of certain GQDs is due to the
presence of unequal number of sublattice atoms in these GQDs, that is, N$_{A}$
- N$_{B}$ $\neq$ 0. Tunability of HLG has also been reflected in the tunability
of the absorption profiles in these GQDs. We find that majority of the GQDs
absorb strongly in the UV-VIS region with their $\beta_{max}$
($\textit{p}_{max}$) being located in the region between 250-450 (300-700) nm)
and their $\alpha$ and $\beta$ values are in the range of 250$-$700 a.u. and
1$-$200 a.u., respectively. However, $\sim$ 70 GQDs have their
$\textit{p}_{max}$ in IR-region and have higher $\alpha$ ($>$ 700 a.u.) and
$\beta$ ($>$ 200 a.u.) values.
A common feature which we find in all these IR-active GQDs is the existence of
larger number of nodes in the isosurface of HOMO which leads to an increment in
HOMO energy, and hence, decrement in the HLG. Due to their high oscillator
strengths in both UV-VIS and IR-regions these GQDs can possess broad band
absorption. With their high $\alpha$ and $\beta$ values along with the BBA, we
expect them to be potential candidates for optoelectronic devices.

\bibliography{nlo_gqd}{}

\section{Supporting Information} Table containing the wavelength, OS and MO
contributions to the $\textit{p}_{max}$ transition for the GQDs with
$\textit{p}_{max}$ in IR-region are given.  Absorption spectra and isosurfaces
of HOMO are given for C32 and C74 GQDs. Comparison of ZINDO/s results with
TDDFT and experimental results are also given. This material is available free
of charge via the Internet at http://pubs.acs.org.

\section{Acknowledgments} S.S.R.K.C.Y., M. M and S.K.P. acknowledge TUE-CMS,
JNCASR for the computational facilities and DST for funding. S.S.R.K.C.Y. thank
Dr. Noel for his help in GaussSum.

\clearpage
\begin{figure}
\centerline{\huge{\bf {Graphical TOC entry}}}
\center\includegraphics[scale=0.4]{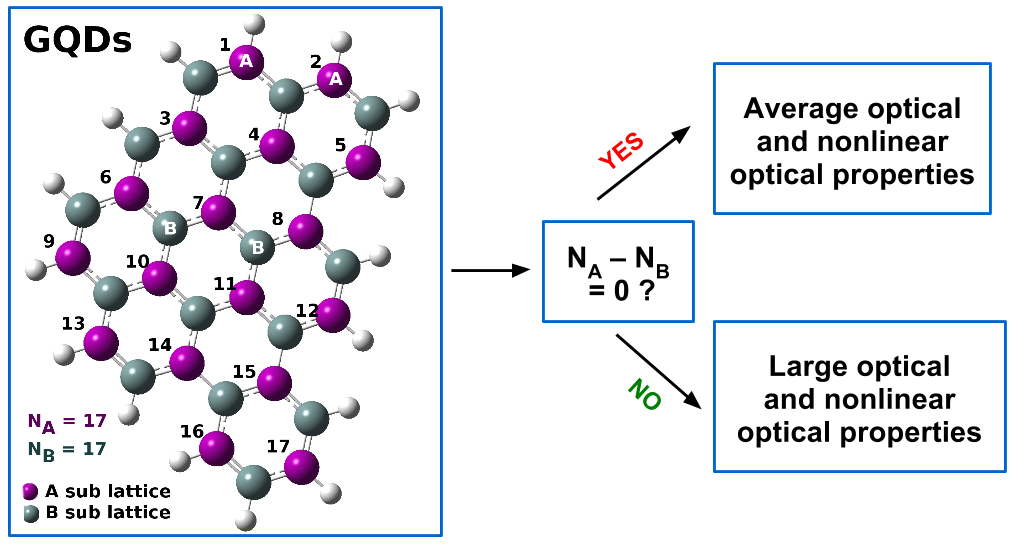}
\caption*{GQDs with nonequivalent number of sublattice atoms show larger linear
and nonlinear optical properties.}
\end{figure}




\begin{table}[h]

\caption*{$\textbf{Table S1:}$Reported values of the absorption, polarizability and 1st
hyperpolarizability (at static field) for p-niitro aniline along with the results from the present
work}
\begin{tabular}{|c|>{\centering\arraybackslash}p{4cm}|>{\centering\arraybackslash}p{4cm}|>{\centering\arraybackslash}p{4cm}|}      \hline  
Methods                   & Absorption (nm)                                 & $\alpha$$_{av}$ (a.u.)  & $\beta$$_{av}$ (a.u.)      \\ \hline 
Experimental              & 350 (in dioxane)$^{a}$ \newline 375 (in water)  & 114.7$^{b}$             & 1072(+/-44)$^{c}$          \\ \hline 
DFT reported              & 291$^{d}$                                       & 102.8$^{e}$             & 1794.7$^{e}$ ($\beta^{x}$) \\ \hline 
QM/MM                     & 266$^{f}$                                       & 110.3$^{g}$             & 978.2$^{g}$                \\ \hline  
DFT$^{h}$                 & 269                                             &  94.9                   & 3741                       \\ \hline 
Semiempirical$^{h}$       & 316                                             & 100.5                   & 492.5                      \\ \hline 
\end{tabular} \\

\begin{flushleft}
\begin{tiny}
a) Ultrafast Charge-Transfer Dynamics: Studies of p-Nitroaniline in Water and Dioxane, C. L.
Thomsen, J. Thøgersen, and S. R. Keiding J. Phys. Chem. A 1998, 102, 1062-1067.\\
b) Experimental Investigations of Organic Molecular Nonlinear Optical Polarizabilities. 1. Methods
and Results on Benzene and Stilbene Derivatives   Ching, L-T.; Tam, W.; Stevenson, S. H.; Meredith,
G.; Rikken, G.; Marder, S. R. J Phys Chem 1991, 95, 10631.\\
c) A comparison of molecular hyperpolarizabilities from gas and liquid phase measurements Kaatz, P.;
Donley, E. A.; Shelton, D. P. J Chem Phys 1998,108, 849.\\
d) Relaxation of Optically Excited p-Nitroaniline: Semiempirical Quantum-Chemical Calculations
Compared to Femtosecond Experimental Results,  Vadim M. Farztdinov, Roland Schanz, Sergey A.
Kovalenko, and Nikolaus P. Ernsting J. Phys. Chem. A 2000, 104, 11486-11496.\\
e) B3LYP Study of the Dipole Moment and the Static Dipole (Hyper)Polarizabilities of
para-Nitroaniline in Gas Phase. Int. J.  Quan. Chem. 2006, 106, 1130–1137.\\
f) Solvent Effects on the Electronic Transitions of p-Nitroaniline: A QM/EFP Study, Dmytro Kosenkov
and Lyudmila V. Slipchenko, J. Phys. Chem. A 2011, 115, 392–401.\\
g)The first hyperpolarizability of p-nitroaniline in 1,4-dioxane: A quantum mechanical/molecular
mechanics study, Lasse Jensen, Piet Th. van Duijnen, J. Chem. Phys. 2005, 123, 074307.\\
h)This work. DFT at CAMB3LYP/6-31+g(d) level of theory. Semiempirical results of absroption are
calculated with ZINDO/S and $\alpha$ and $\beta$ with MOPAC.\\
\end{tiny}
\end{flushleft}

\end{table}

\begin{figure}[h] 
\center\includegraphics[scale=0.2]{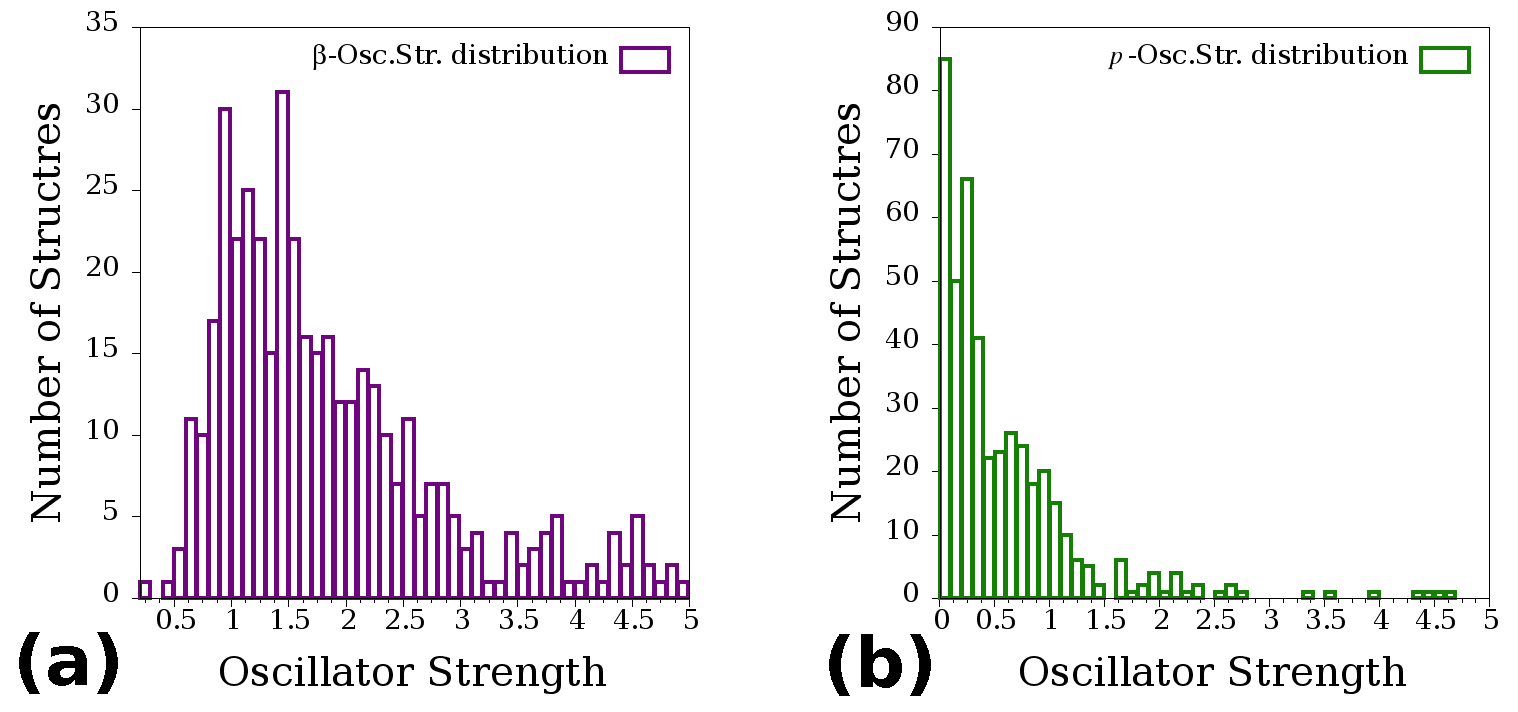}
\caption*{$\textbf{Figure S1:}$ Histograms of oscillator strengths corresponding to (a) $\beta_{max}$, (b) $\textit{p}_{max}$
excitation in all GQDs.} 
\end{figure}

\clearpage
\begin{longtable}{|c|c|c|>{\centering\arraybackslash}p{8cm}|}

Systems  & Wavelength (nm) & Osc. Str. &   MO contributions  \\ \hline
c38hf	 &  764.6684 & 0.9090 & HOMO$\rightarrow$LUMO (93\%) \\ \hline
c42hd	 &  781.5389 & 0.2922 & HOMO$\rightarrow$LUMO (93\%) \\ \hline
c74ha	 &  784.6053 & 9.4406 & HOMO$\rightarrow$L+1 (45\%), H-1$\rightarrow$LUMO (46\%) \\ \hline
c60hm	 &  799.0677 & 1.6014 & HOMO$\rightarrow$LUMO (91\%) \\ \hline
c54he	 &  803.7818 & 0.8854 & HOMO$\rightarrow$LUMO (91\%) \\ \hline
c78h42	 &  805.0343 & 0.5534 & H-1$\rightarrow$L+1 (69\%), H-2$\rightarrow$L+2 (22\%) \\ \hline
c38ha	 &  808.7628 & 0.3008 & HOMO$\rightarrow$LUMO (93\%) \\ \hline
c60hk	 &  812.7390 & 1.1520 & HOMO$\rightarrow$LUMO (93\%) \\ \hline
c54hh	 &  823.9738 & 0.6357 & HOMO$\rightarrow$LUMO (93\%) \\ \hline
c82h44	 &  827.3278 & 0.5781 & H-1$\rightarrow$L+1 (66\%), H-2$\rightarrow$L+2 (25\%) \\ \hline
c42hh	 &  827.8249 & 0.9441 & HOMO$\rightarrow$LUMO (93\%) \\ \hline
c50hd	 &  856.1794 & 0.9842 & HOMO$\rightarrow$LUMO (93\%) \\ \hline
c54hc	 &  859.8012 & 0.3036 & HOMO$\rightarrow$LUMO (91\%) \\ \hline
c42hc	 &  860.1591 & 0.3094 & HOMO$\rightarrow$LUMO (92\%) \\ \hline
c50hb	 &  882.9464 & 0.3115 & HOMO$\rightarrow$LUMO (91\%) \\ \hline
c60hb	 &  893.8313 & 0.7016 & HOMO$\rightarrow$LUMO (82\%) \\ \hline
c54hg	 &  900.3873 & 0.9397 & HOMO$\rightarrow$LUMO (93\%) \\ \hline
c60ho	 &  902.2876 & 1.7710 & HOMO$\rightarrow$LUMO (94\%) \\ \hline
c60hf	 &  904.4597 & 0.6102 & HOMO$\rightarrow$LUMO (92\%) \\ \hline
c74hs	 &  905.2522 & 0.0556 & HOMO$\rightarrow$L+1 (72\%), HOMO$\rightarrow$L+5 (11\%) \\ \hline
c60hh	 &  908.3694 & 1.0712 & HOMO$\rightarrow$LUMO (92\%) \\ \hline
c60hp	 &  910.4372 & 1.8297 & HOMO$\rightarrow$LUMO (94\%) \\ \hline
c60hr	 &  920.8507 & 1.8023 & HOMO$\rightarrow$LUMO (94\%) \\ \hline
c54hb	 &  923.8699 & 0.3222 & HOMO$\rightarrow$LUMO (90\%) \\ \hline
c74hr	 &  928.7141 & 1.1620 & HOMO$\rightarrow$LUMO (93\%) \\ \hline
c60ha	 &  941.0500 & 0.3354 & HOMO$\rightarrow$LUMO (88\%) \\ \hline
c50ha	 &  942.9108 & 0.3294 & HOMO$\rightarrow$LUMO (89\%) \\ \hline
c60he	 &  959.4006 & 0.9232 & HOMO$\rightarrow$LUMO (92\%) \\ \hline
c60hg	 &  959.8462 & 0.6174 & HOMO$\rightarrow$LUMO (93\%) \\ \hline
c54ha	 &  975.8625 & 0.3409 & HOMO$\rightarrow$LUMO (88\%) \\ \hline
c54hf	 &  985.0110 & 0.9331 & HOMO$\rightarrow$LUMO (92\%) \\ \hline
c74he	 &  994.1732 & 1.0510 & HOMO$\rightarrow$LUMO (87\%) \\ \hline
c58h32	 & 1002.9391 & 0.3537 & HOMO$\rightarrow$LUMO (87\%) \\ \hline
c74hd	 & 1010.7063 & 1.0094 & HOMO$\rightarrow$LUMO (82\%) \\ \hline
c74hc	 & 1024.2324 & 0.3657 & HOMO$\rightarrow$LUMO (85\%) \\ \hline
c62h34	 & 1026.2672 & 0.3674 & HOMO$\rightarrow$LUMO (85\%) \\ \hline
c66h36	 & 1045.7434 & 0.3821 & HOMO$\rightarrow$LUMO (84\%) \\ \hline
c74hb	 & 1053.2054 & 0.3891 & HOMO$\rightarrow$LUMO (82\%) \\ \hline
c70h38	 & 1062.0467 & 0.3977 & HOMO$\rightarrow$LUMO (82\%) \\ \hline
c60hc	 & 1064.8745 & 1.0999 & HOMO$\rightarrow$LUMO (92\%) \\ \hline
c74hg	 & 1191.0023 & 1.1662 & HOMO$\rightarrow$LUMO (92\%) \\ \hline
c60hd	 & 1193.2949 & 0.9702 & HOMO$\rightarrow$LUMO (91\%) \\ \hline
c74hf	 & 1209.8296 & 1.2112 & HOMO$\rightarrow$LUMO (91\%) \\ \hline
c74hh	 & 1359.3174 & 1.0917 & HOMO$\rightarrow$LUMO (91\%) \\ \hline
c74hi	 & 1405.2288 & 0.9118 & HOMO$\rightarrow$LUMO (89\%) \\ \hline
c74hw	 & 1793.4809 & 1.0937 & HOMO$\rightarrow$LUMO (90\%) \\ \hline
c74hz	 & 1794.7790 & 1.1039 & HOMO$\rightarrow$LUMO (90\%) \\ \hline
c74hy	 & 1911.5531 & 1.0413 & HOMO$\rightarrow$LUMO (87\%) \\ \hline
c74hx	 & 1927.6016 & 1.0427 & HOMO$\rightarrow$LUMO (87\%) \\ \hline
c33h15	 & 1980.5645 & 0.0213 & HOMO(B)$\rightarrow$L+1(B) (61\%), H-1(A)$\rightarrow$LUMO(A) (13\%) \\ \hline
c74hp	 & 2049.6501 & 0.9875 & HOMO$\rightarrow$LUMO (84\%) \\ \hline
c74ho	 & 2068.8025 & 0.9320 & HOMO$\rightarrow$LUMO (84\%) \\ \hline
c74hn	 & 2104.2657 & 0.8585 & HOMO$\rightarrow$LUMO (83\%) \\ \hline
c74hm	 & 2176.2917 & 0.7512 & HOMO$\rightarrow$LUMO (83\%), H-1$\rightarrow$L+1 (10\%) \\ \hline
c78h24	 & 2427.7137 & 0.2894 & HOMO$\rightarrow$LUMO (14\%), HOMO$\rightarrow$L+1 (20\%), H-1$\rightarrow$LUMO (20\%), H-1$\rightarrow$L+1 (14\%), H-1$\rightarrow$L+2 (21\%) \\ \hline
c28hcc	 & 3158.0065 & 0.2001 & HOMO$\rightarrow$LUMO (87\%) \\ \hline
c46h18	 & 4221.4279 & 0.1481 & HOMO$\rightarrow$LUMO (50\%), H-1$\rightarrow$L+1 (36\%) \\ \hline
c28hS	 & 4251.8291 & 0.1646 & HOMO$\rightarrow$LUMO (86\%) \\ \hline
c24h14c	 & 4335.0817 & 0.0951 & HOMO$\rightarrow$LUMO (90\%) \\ \hline
c24hc	 & 4335.0817 & 0.0951 & HOMO$\rightarrow$LUMO (90\%) \\ \hline
c32he	 & 4359.4703 & 0.1278 & HOMO$\rightarrow$LUMO (90\%) \\ \hline
c28hM	 & 4413.7891 & 0.1169 & HOMO$\rightarrow$LUMO (88\%) \\ \hline
c32hc	 & 4659.2761 & 0.1125 & HOMO$\rightarrow$LUMO (90\%) \\ \hline
c28hR	 & 6156.0743 & 0.0708 & HOMO$\rightarrow$LUMO (89\%) \\ \hline
c74hj	 & 6515.1517 & 0.0958 & HOMO$\rightarrow$LUMO (58\%), H-1$\rightarrow$LUMO (34\%) \\ \hline
c28hss	 & 6687.3429 & 0.0515 & HOMO$\rightarrow$LUMO (91\%) \\ \hline
c38hk	 & 7505.0446 & 0.0535 & HOMO$\rightarrow$LUMO (89\%) \\ \hline
c28hj	 &10818.7903 & 0.0297 & HOMO$\rightarrow$LUMO (92\%) \\ \hline
c30hj	 &12167.1577 & 0.0236 & HOMO$\rightarrow$LUMO (92\%) \\ \hline
c34hg	 &12251.3178 & 0.0284 & HOMO$\rightarrow$LUMO (92\%) \\ \hline
c26hR	 &18126.2188 & 0.0137 & HOMO$\rightarrow$LUMO (93\%) \\ \hline

\caption*{$\textbf{Table S2:}$ System names, wavelength corresponding to, oscillator strength (Osc.
Str.) of and molecular orbital contributions for ``$\textit{p}_{max}$ excitation" of all the GQDs
whose $\textit{p}_{max}$ is in IR-region. Names of the GQDs are as given in Ref 23 of the main
article. All the corresponding structures can be obtained from this link.
http://journals.aps.org/prb/supplemental/10.1103/PhysRevB.81.085430/GNFs\_PAHs\_coord.tar.gz} 

\end{longtable}


\begin{figure}[h] 
\center\includegraphics[scale=0.1]{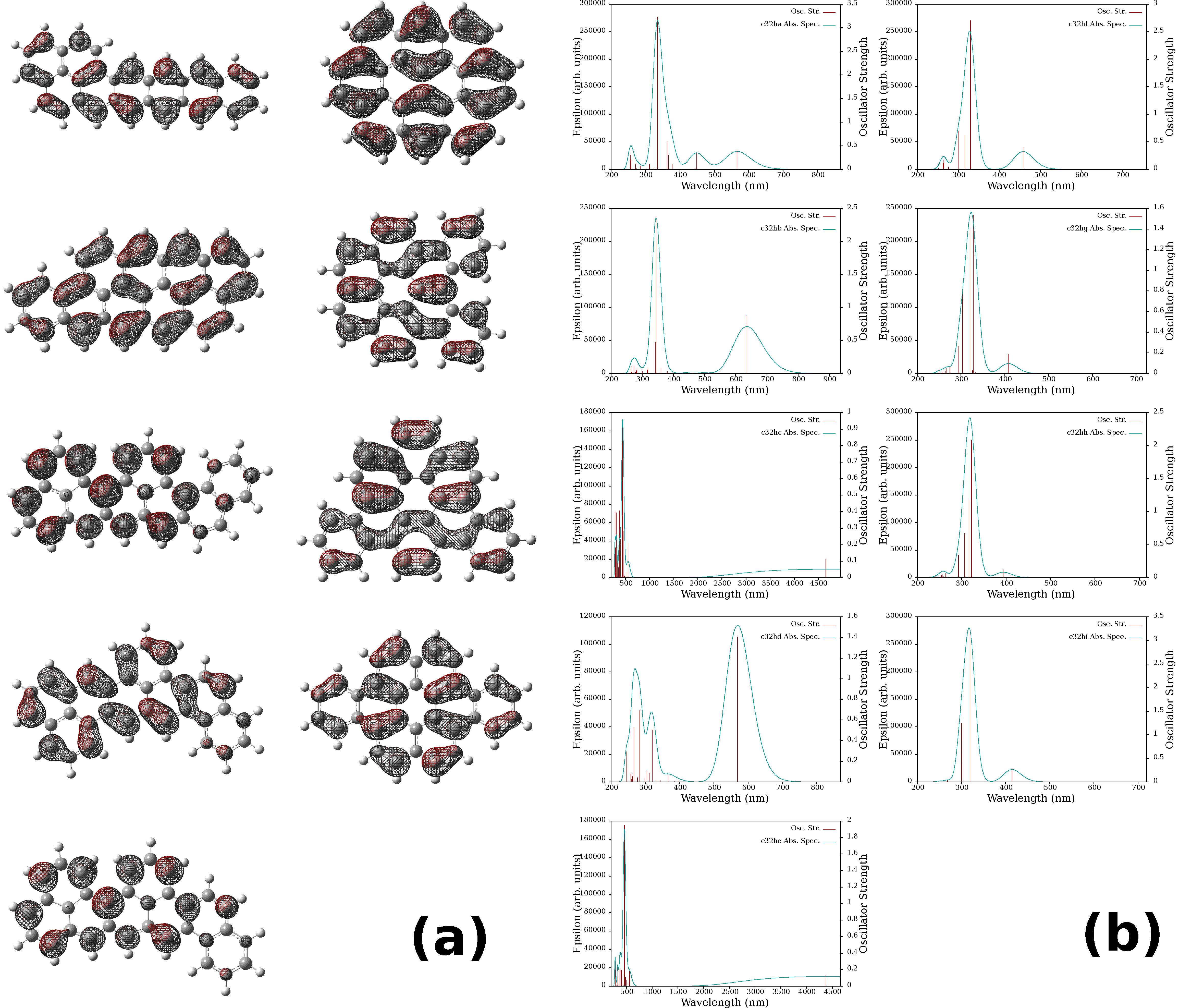}
\caption*{$\textbf{Figure S2:}$ Panel (a) shows the isosurfaces of HOMO of all C32 GQDs considered in
this study. Panel (b) shows their corresponding absorption spectra.}
\end{figure}

\begin{figure}[h] 
\center\includegraphics[scale=0.15]{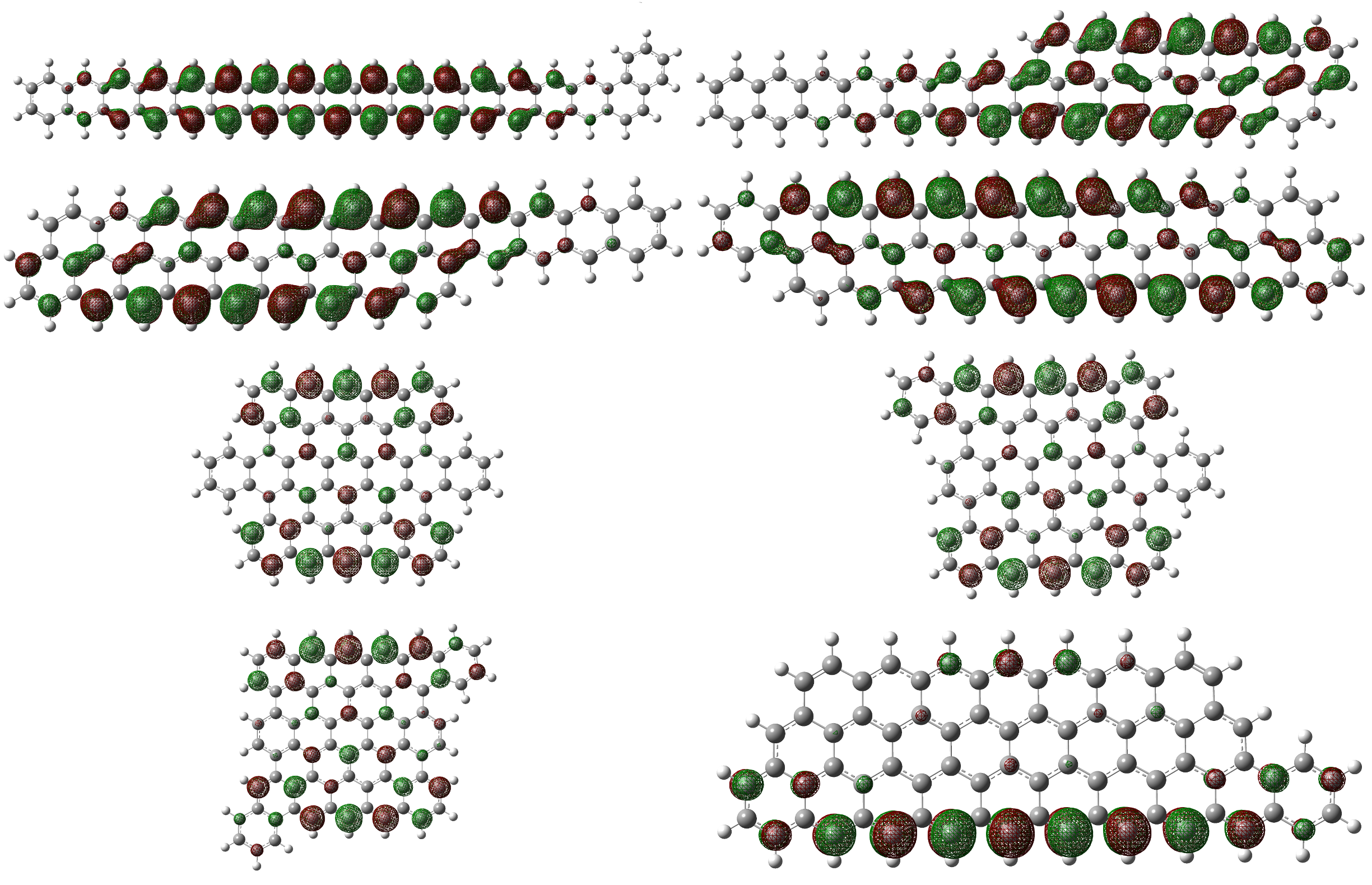}
\caption*{$\textbf{Figure S3:}$ Isosurfaces of HOMO of some of the C74 GQDs considered in
this study.} 
\end{figure}

\begin{figure}[h] 
\center\includegraphics[scale=0.15]{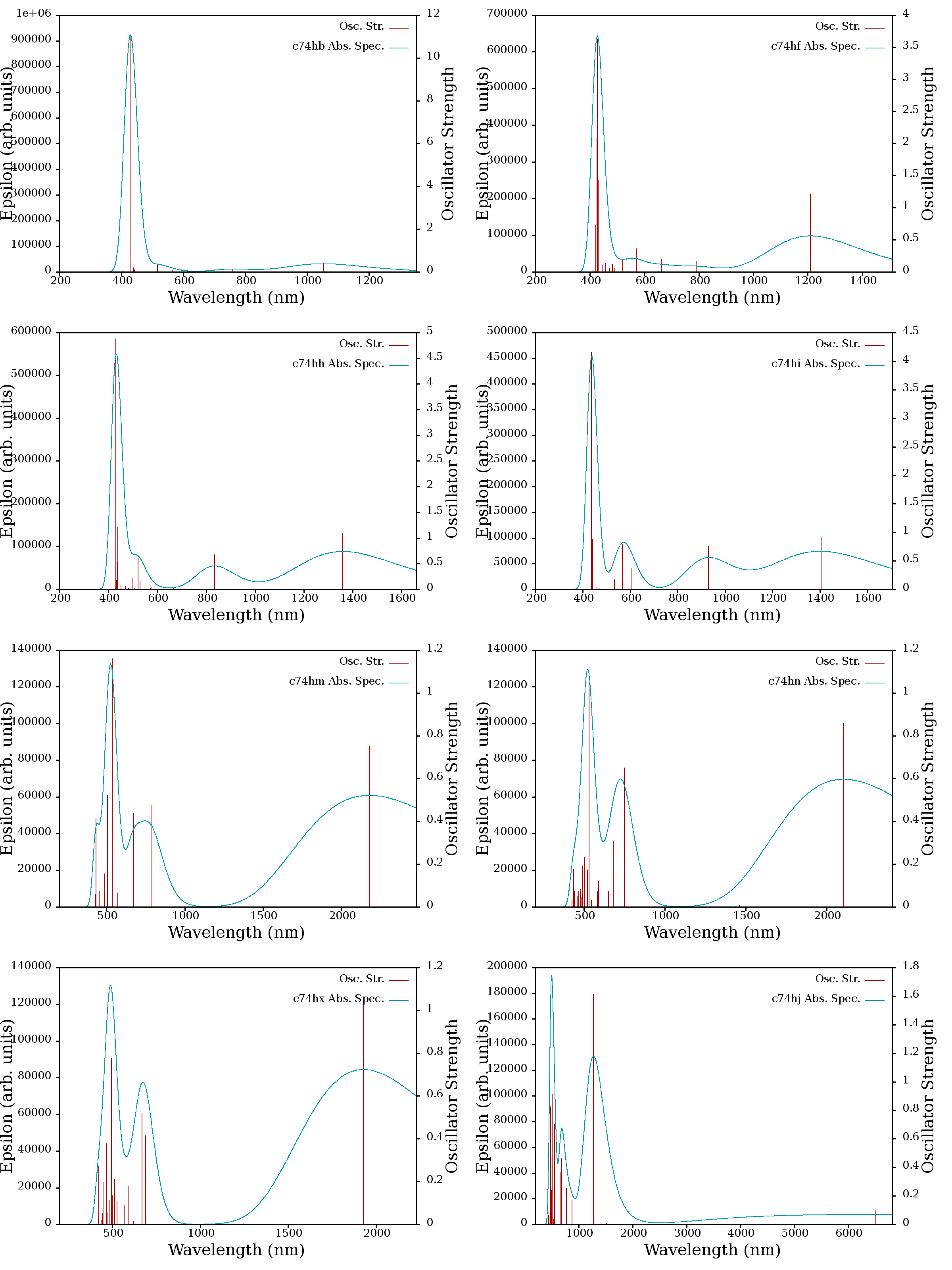}
\caption*{$\textbf{Figure S4:}$ Absorption spectra corresponding to the GQDs given in the above figure (Fig. S3).} 
\end{figure}


\begin{table}[h]
\caption*{$\textbf{Table S3:}$ System names, wavelength corresponding to ``$\textit{p}_{max}$" and "$\beta_{max}$"
excitations of all C28-GQDs whose $\textit{p}_{max}$ is in IR-region are given. Values inside the
paranthesis are the ZINDO/S results and the ones which are outside are the CAM-B3LYP/6-31+g(d)
results.}
\begin{tabular}{|c|c|c|} \hline
System & $\textit{p}_{max}$ (nm) &  $\beta_{max}$ (nm)  \\ \hline
c28hcc &  2966.1085 ( 3158.0065) &  436.1311 (460.6820) \\ \hline
c28hj  &  4125.9014 (10818.7903) &  346.5642 (429.1269) \\ \hline
c28hM  &  3035.8310 ( 4413.7891) &  434.3282 (460.0666) \\ \hline
c28hR  &  3145.1887 ( 6156.0743) &  393.7104 (454.2180) \\ \hline
c28hS  &  2896.1303 ( 4251.8291) &  425.8401 (491.0231) \\ \hline
c28hss &  3408.0082 ( 6687.3429) &  397.7012 (448.3378) \\ \hline
\end{tabular}
\end{table}

\begin{table}[h]
\caption*{$\textbf{Table S4:}$ System name, polarizability and 1st hyperpolarizability of C28-GQDs
Values inside the paranthesis are the MOPAC results and the outside ones are CAM-B3LYP/6-31+g(d)
results.}
\begin{tabular}{|c|c|c|} \hline
System &   $\alpha$ (a.u)                &        $\beta$ (a.u)                   \\ \hline
c28hcc &   744.480   (1502.36167)        &       97762.095        (485305.1890)    \\ \hline
c28hM  &   761.495   (1696.01550)        &      199892.814       (1304022.4046)    \\ \hline
c28hss &   801.149   (1726.62126)        &      437976.831       (2003517.2092)    \\ \hline
c28hj  &   822.574   (2264.26242)        &      869985.316       (5567249.5402)    \\ \hline
\end{tabular}
\end{table}

\end{document}